\documentclass[twocolumn]{aastex631}
\usepackage{CJK}
\shorttitle{4U 1820-30}
\shortauthors{Chou et al.}

\begin{document}
\begin{CJK*}{Bg5}{bsmi}
\title{The Puzzling Superorbital Period Variation of the Low-mass X-ray Binary 4U 1820-30}

\correspondingauthor{Yi Chou}
\email{yichou@astro.ncu.edu.tw}

\author[0000-0002-8584-2092]{Yi Chou (©PÖö)}
\affiliation{Graduate Institute of Astronomy, National Central University \\
  300 Jhongda Rd. Jhongli Dist. Tauyuan, 320317, Taiwan}
\author{Jun-Lei Wu (§d§g½U)}
\affiliation{Graduate Institute of Astronomy, National Central University \\
  300 Jhongda Rd. Jhongli Dist. Tauyuan, 320317, Taiwan}
\author{Bo-Chun Chen (³¯ªyëÞ)}
\affiliation{Graduate Institute of Astronomy, National Central University \\
  300 Jhongda Rd. Jhongli Dist. Tauyuan, 320317, Taiwan}
\author{Wei-Yun Chang (±iÞ³ªå)}
\affiliation{Graduate Institute of Astronomy, National Central University \\
  300 Jhongda Rd. Jhongli Dist. Tauyuan, 320317, Taiwan}



\begin{abstract}

Because of the previously observed stability of the 171-day period, the superorbital modulation of the low-mass X-ray binary 4U 1820-30 was considered a consequence of a third star orbiting around the binary. This study aims to further verify this triple model by testing the stability of superorbital period using the light curves collected by X-ray sky monitoring/scanning telescopes from 1987 to 2023. Both power spectral and phase analysis results indicate a significant change in the superorbital period from 171 days to 167 days over this 36-year span. The evolution of the superorbital phase suggests that the superorbital period may have experienced an abrupt change between early 2001 and mid-2003 or changed gradually with a period derivative of  $\dot P_{sup}=(-4.20 \pm 0.72) \times 10^{-4}$ day/day. We conclude that the superorbital period of 4U 1820-30 was not as stable as anticipated by the triple model, which strongly challenges this hypothesis. Instead, we propose an irradiation-induced mass transfer instability scenario to explain the superorbital modulation of 4U 1820-30.

\end{abstract}



\section{Introduction} \label{intro}

4U 1820-30, discovered by~\citet{gia74}, is an ultra-compact low mass X-ray binary (LMXB) located near the center of globular cluster NGC 6624. It was the first X-ray source known to exhibit Type-I X-ray burst~\citep{gri76}, indicating that the accretor in this binary system is a neutron star. Its 685 s orbital period, discovered by~\citet{ste87} from its sinusoidal-like orbital modulation in the X-ray light curve, makes 4U 1820-30 the most compact LMXB. The mass-losing companion is a Roche-lobe filling helium white dwarf with a mass of 0.06-0.08 $M_{\sun}$~\citep{rap87}. Mass transfer in the system is induced by the orbital angular momentum loss through gravitational radiation, which should result in a positive orbital period derivative with a lower limit of $\dot P_{orb}/P_{orb} > 8.8 \times 10^{-8}$ $yr^{-1}$~\citep{rap87}. However, observed orbital period derivatives reported by~\citet{tan91,van93a,van93b},~\citet{chou01} (hereafter CG01), and~\citet{peu14} were negative with the latest value of  $\dot P_{orb}/P_{orb} = (-5.21 \pm 0.13) \times 10^{-8}$ $yr^{-1}$ updated by~\citet{chou23}, evaluated from $\sim$46 years of orbital phase evolution. This contradiction is believed due to the binary system accelerating by the gravitational potential in NGC 6624~\citep{tan91,chou01,peu14,chou23}. Additionally, superhump modulation with a period of $691.6\pm 0.7$ s, $\sim$1\% significantly longer than the orbital period, was observed in both FUV~\citep{wan10} and X-ray~\citep{chou23} bands. From the superhump period, the mass of companion of 4U 1820-30 was estimated as 0.07$M_{\sun}$~\citep{wan10,chou23}. 

In addition to orbital and superhump variations, 4U 1820-30 exhibits superorbital modulation with a period much longer than its orbital period.~\citet{pri84} discovered the X-ray flux modulation by a factor of 2 with a period of $176.4 \pm 1.3$ days using the light curve detected by Vela 5B from 1969 to 1976. This periodicity was further confirmed by~\citet{sm92}. However, by analyzing the light curve collected between 1996 and 2000 by All Sky Monitor on-board Rossi X-ray Timing Explorer (RXTE ASM), CG01 revised the superorbital period to  $171.39 \pm 1.93$ days. Combining the times of the flux minima of the superorbital modulation (hereafter superorbital minima) detected by Vela 5B and All Sky Monitor onboard Ginga (Ginga ASM), CG01 further constrained the period to be $171.033 \pm 0.326$ days and claimed that the superorbital period was stable over $\sim$30 years with $|\dot P_{sup} / P_{sup} | < 2.2 \times 10^{-4}$ $yr^{-1}$. Based on the stability of the superorbital period, CG01 proposed that this long-term variability is due to a hierarchical third star orbiting around the binary system~\citep[][hereafter the triple model]{gri86,gri88}. The hierarchical third component can cause the eccentricity of inner binary system to vary with a period ($P_{ecc}$) as

\begin{equation}\label{msp}
 P_{ecc}= K {{P_3}^2 \over P_{orb}}
\end{equation}

\noindent where $P_3$ is the orbital period of third star, $P_{orb}$ is the binary orbital period and $K$ is a constant of unity~\citep{ma79}. Because the mass transfer rate is highly sensitive to the Roche lobe radius, which is proportional to the binary separation, the variation of binary eccentricity can cause the mass loss rate and the accretion rate to change with a period of $P_{ecc}$ and thus $P_{sup}=P_{ecc}$. For the 4U 1820-30 system, the orbital period of the third companion is estimated to be $\sim$1.1 days for $K \sim 1$, and beat sidebands resulting from coupling binary modulation and $\sim$1.1 day periodicity may be observable in the power spectrum. Although these beat sidebands were not detected in RXTE observations (CG01),~\citet{chou23} suggested that the $691.6\pm 0.7$ s periodicity observed in the X-ray band might be caused by a hierarchical triple orbiting around the binary system with an orbital period of 0.8 days. Moreover, CG01 found that the active times of Type-I X-ray bursts were clustered within $\pm$23 days of expected superorbital minima, which aligns with the observation that the bursts can be seen only in low state~\citep{cla77,ste84}. This fact implies that the superorbital modulation of 4U 1820-30 is due to changes in the accretion rate rather than external occultation or absorption effects, which is consistent with the triple model. 

The periodicity of 171 days was further confirmed by~\citet{sim03,wen06,zdz07a,kot12} using additional RXTE ASM data and by~\citet{far09} using the data collected by Burst Alert Telescope onboard the Neil Gehrels Swift Observatory (Swift BAT). Applying the triple model,~\citet{zdz07a} demonstrated that the factor of 2 superorbital modulation in X-ray light curve can be explained by the eccentricity of inner binary oscillating between 0 and 0.004. The discovery of the dependence of orbital modulation profile on the accretion rate~\citep{zdz07b} also supports the triple model. The hard X-ray light curve collected from Swift BAT showed that the superorbital modulation can be observable only for the energy bands less than 24 keV~\citep{far09}. Conversely, by comparing the peak widths of the power spectra made from light curves detected by RXTE ASM and Swift BAT with the corresponding simulated light curves,~\citet{far09} found that the peak widths from real data are marginally wider than the ones from simulated data, concluding that this may be caused by the superorbital period change.~\citet{kot12} adopted the dynamic power spectrum technique to analyze the superorbital variability of several X-ray binaries, and found no significant superorbital period change for 4U 1820-30 except for a weakening of power during MJD $\sim$51,200-52,200.

Owing to the monitoring/scanning X-ray telescopes, 4U 1820-30 has been observed for decades and is still being monitored by the Swift BAT the Monitor of All-sky X-ray Image (MAXI), and Fermi Gamma-ray Burst Monitor (Fermi GBM). In this work, we aim to further verify the stability of the superorbital period, which is the crucial evidence for the triple model of 4U 1820-30 system, and to establish an updated ephemeris for superorbital modulation. In this paper, we introduce the instruments used to obtain the light curves for this research, including Ginga ASM, RXTE ASM, Swift BAT, MAXI, and Fermi GBM as well as the light curve collected by RXTE Proportional Counter Array (RXTE PCA) while it processed the monitoring observations of the Galactic center and plane~\citep{mar06}, in Section~\ref{obs}. A preliminary superorbital period stability test was performed using the power spectrum made by the entire light curve of each instrument (Section~\ref{ps}). A more detailed measurement of superorbital period variation was obtained by analyzing the superorbital phase evolution and updating the ephemerides (Section~\ref{pha}). The new ephemerides allow us to verify whether the Type-I X-ray bursts occur clustered around the expected superorbital minima. In Section~\ref{dis}, we discuss the instability of superorbital period, which poses a serious challenge of the triple model, and explore the possible interpretations for the superorbital period variation of 4U 1820-30.


\section{Observations} \label{obs}
The Ginga ASM consisted of two identical gas proportional counters with six fan-beam collimators to restrict field of view (FOV) of $1^{\circ} \times 45^{\circ}$. It was sensitive to X-ray photons with energies between 1 and 20 keV, and  had a total effective area of 420 cm$^2$. It provided real-time alerts of X-ray transient phenomena and long-term historical records of X-ray sources. The Ginga ASM monitored the sky from 1987 February to 1991 October. Further details of the Ginga ASM are described by~\citet{tsu89}. The Ginga ASM light curve of 4U 1820-30 archived on the website of the High Energy Astrophysics Science Archive Research Center (HEASARC\footnote{https://heasarc.gsfc.nasa.gov/docs/archive.html}) collected from March 7, 1987 (MJD 46,861) to October 3, 1991 (MJD 48,532) was analyzed in this study.

The RXTE ASM~\citep{lev96} was an instrument mounted on RXTE to monitor the variable and the transient X-ray sources. It consisted of three scanning shadow cameras, each containing a position-sensitive proportional counter, to observe the sky through a one-dimensional coded mask with an FOV of $6^{\circ} \times 90^{\circ}$. It was designed to detect the cosmic X-rays in the photon energy range of 1.5-12 keV, which can be further divided into 1.5-3, 3-5 and 5-12 keV energy bands. In addition to these energy bands, the light curves with two different time resolutions, dwell (a 90 sec exposure) and one-day binned, were also archived. During its mission, from the beginning of 1996 to early 2012, the RXTE ASM scanned the entire sky every 90 minutes. In this research, the 1.5-12 keV RXTE ASM light curve of 4U 1820-30 archived on the website of the HEASARC collected between January 6, 1996 (MJD 50,088) and September 27, 2011 (MJD 55,831) was selected to analyze the superorbital modulation of 4U 1820-30.

In addition to the regular pointing observations, the RXTE PCA also conducted monitoring observations of the galactic center and plane starting from 1999~\citep{mar06}. The PCA was an instrument with an effective area of 6500 cm$^2$ designed to detect the X-ray photons in the energy range of 2-60 keV~\citep{jah96}. Despite being a non-imaging instrument, its $1^{\circ}$ FOV, constrained by collimators, allowed for identification and detection of X-ray sources. It scanned over galactic bulge and plane approximately twice per week~\citep{mar06}, providing sufficient cadence to resolve the superorbital modulation of 4U 1820-30. The light curve of 4U 1820-30 collected by PCA in this program from February 5, 1999 (MJD 51,214) to October 30 2011 (MJD 55,846) was available on the program website\footnote{https://asd.gsfc.nasa.gov/Craig.Markwardt//galscan/html/4U\_1820-30.html}.

The BAT, an instrument on Swift, is a coded-mask telescope with a large FOV (1.4 steradian) to monitor the hard X-ray sky in the energy range 15-150 keV since 2004~\citep{bar05}. Apart from triggering alerts for gamma-ray bursts, its angular resolution ($\sim$20') and large photon collecting area (5200 cm$^2$) enable monitoring of the known cosmic X-ray sources as the Swift satellite orbits around the Earth every $\sim$96 minutes. This capability allows for the study of long-term variability these sources. In this work, we analyzed the daily binned light curve of 4U 1820-30 that archived on the Swift website\footnote{https://swift.gsfc.nasa.gov/results/transients/H1820-303/} observed from February 14, 2005 (MJD 53,415) through August 1, 2023 (MJD 60,157). 

The MAXI, installed on the Japanese Experiment Module of International Space Station (ISS), is designed to alert the transient X-ray sources and monitor the long-term variations of the X-ray sources~\citep{mat09}. It contains two types of slit cameras with two different detectors: a gas proportional counter with an effective area of 5250 cm$^2$ for detecting the X-ray photons in the energy range of 2-30 keV, and a solid state camera with an effective area of 200 cm$^2$ sensitive to the X-ray photons in the energy range of 0.5-12 keV. MAXI can scan almost the entire sky twice during each ISS orbit ($\sim$90 minutes).  In this study, we analyzed the daily binned light curve of the energy range of 2-20 keV collected between 2009 August 12 (MJD 55,055) and 2023 August 1 (MJD 60,157), available on the MAXI on-demand process website\footnote{http://maxi.riken.jp/mxondem}, to study the superorbital modulation of 4U 1820-30.  

The Fermi GBM is an all-sky monitoring instrument onboard the Fermi Gamma-ray Space Telescope, launched in 2008~\citep{mee09}. Designed to observe gamma-ray bursts and other transient phenomena using the Earth occultation technique~\citep{wil12}, the GBM covers a broad energy range from 8 keV to 40 MeV. It consists of 12 sodium iodide (NaI) detectors, optimized for the low-energy range, and two bismuth germanate (BGO) detectors for higher energies, strategically placed to provide nearly full-sky coverage. The GBM enables rapid localization and characterization of transient gamma-ray events, supporting follow-up observations by other instruments and ground-based facilities. The Fermi GBM light curev of 4U 1820-30 is available on the website of Gamma-Ray Astrophysics Team, National Space, Science, and Technology Center (NSSTC)\footnote{https://gammaray.nsstc.nasa.gov/gbm/science/earth\_occ/H1820-303.html}. In this work, we analyzed the daily binned light curve in the energy range of 12-25 keV collected between 2012 February 9 (MJD 55966), the earliest record in the archived light curve, and 2023 August 1 (MJD 60,157).

\begin{figure}
  \epsscale{1.25}
\plotone{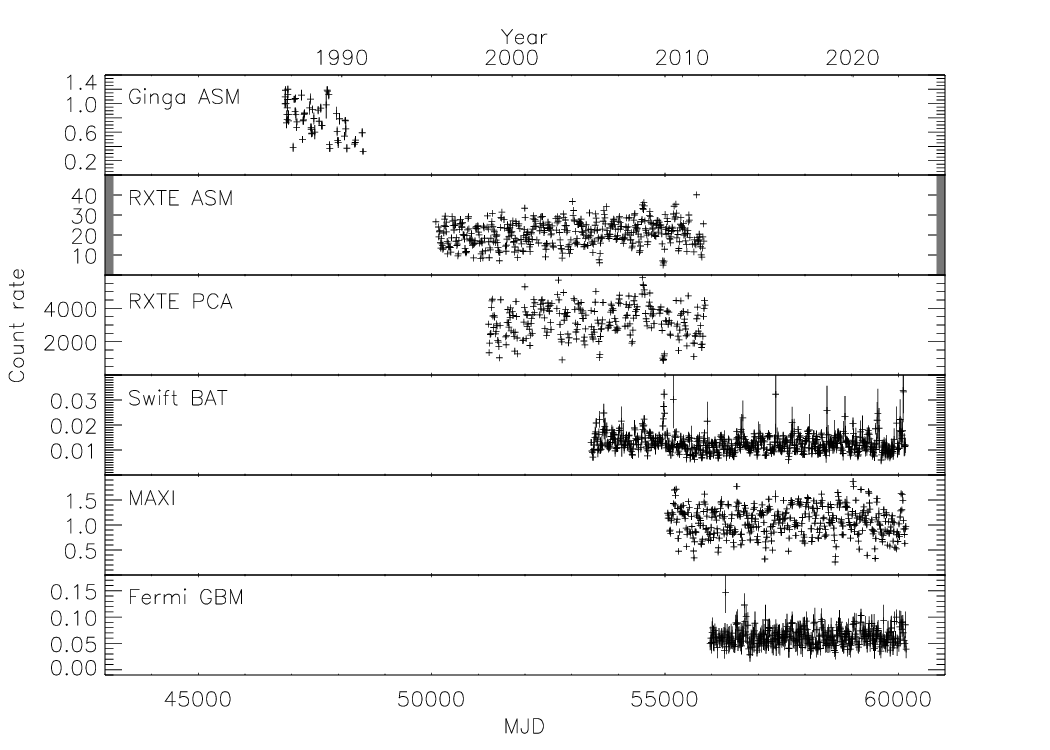}
\caption{Light curves collected by six instruments for analysis in this work. The bin size of these light curves is 10 days.  \label{lcurves}}
\end{figure}

The light curves of 4U 1820-30 collected by these six instruments are shown in Figure~\ref{lcurves}.

\section{Data Analysis} \label{da}
\subsection{Power Spectral Analysis} \label{ps}
In the power spectral analysis, all the light curves were rebinned into daily averages for consistency. To probe the superorbital periods of various observations, the Lomb-Scargle (LS) periodogram~\citep{sca82} was applied to generate the power spectra. We set the frequency resolution to $2 \times 10^{-6}$ cycles/day and then applied the cubic spline interpolation around the peaks of the power spectra to further refine their locations. The errors of signal frequencies were estimated by the method proposed by~\citet{hor86}:

\begin{equation}\label{perr}
 \delta f={{3 \sigma_{N}} \over {4 N_0^{1/2} TA}}
\end{equation}
\noindent where $A$ is the amplitude of the signal, $\sigma_N^2$ is the variance of the noise after the signal being removed, $T$ is the time span of the light curve and $N_0$ is the number of data points. If both signals of the fundamental and second harmonics frequencies are significantly detected, the amplitudes of these components were determined by fitting a two-component sinusoidal function to the light curve, with the frequencies fixed at the signal frequencies derived from the power spectrum. Otherwise, a single sinusoidal function was fitted. $\sigma_N^2$ was estimated as the root-mean-square (rms) of the residuals after subtracting the best-fitting sinusoidal function from the light curve.

\begin{figure}
   \epsscale{1.25}
\plotone{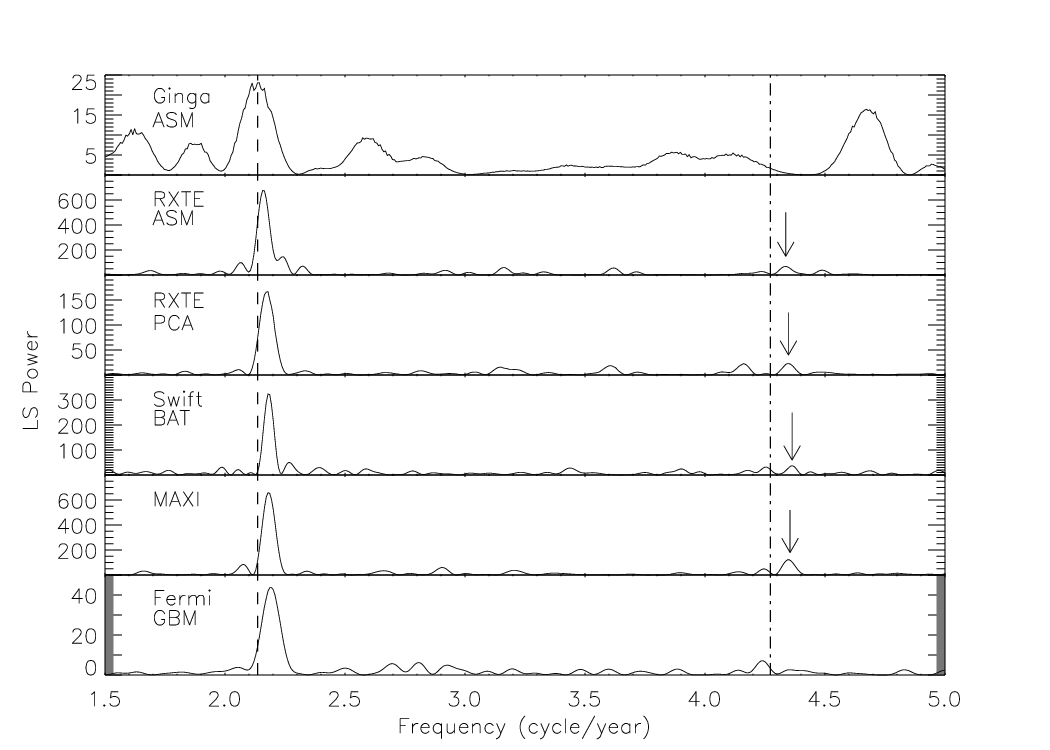}
\caption{Power spectra derived from the light curves collected by six instruments in this work. Significant signals corresponding to the fudamental frequency of the superorbital modulations are visible, while signals of the second harmonic, indicated by arrows, are also detectable with higher-sensitivity instruments. The vertical dashed line marks the superorbital frequency from the CG01 ephemeris (f = 2.136 cycles/year), and the dashed-dotted line represents its second harmonic (f = 4.272 cycles/year). Except for the power spectrum from the Ginga ASM data, notable deviations in the superorbital frequency compared to the one proposed by CG01 can be observed in both the first and second harmonic peaks in the power spectra. \label{powspec}}
\end{figure}

\begin{figure}
  \epsscale{1.2}
  \plotone{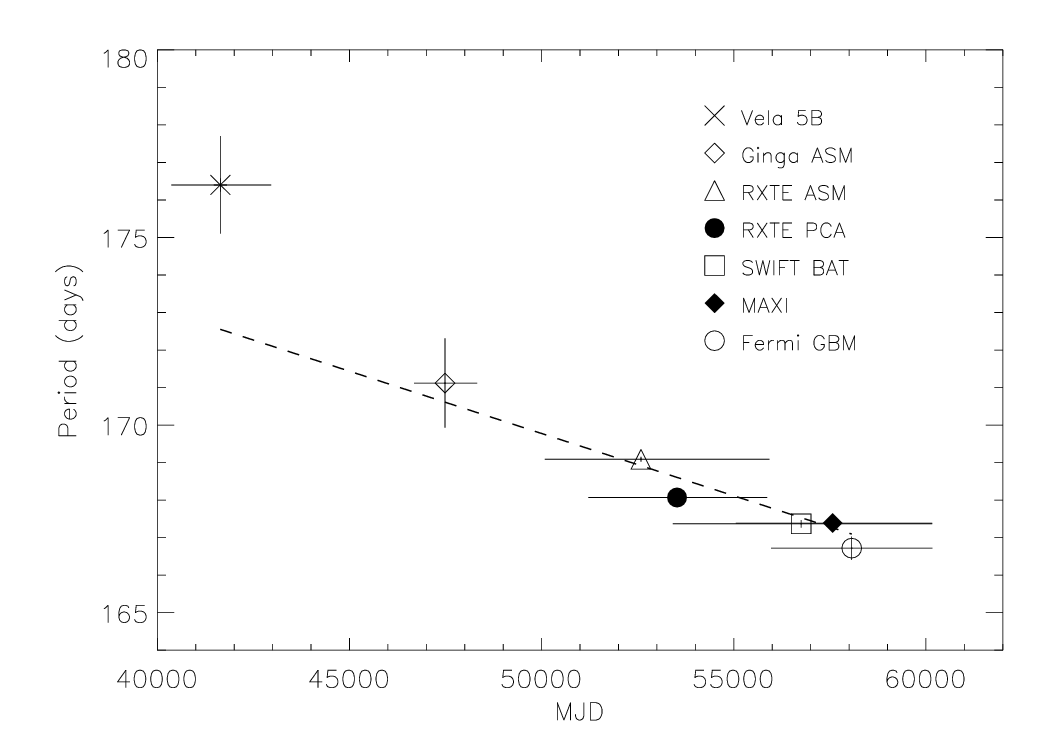}
\caption{Superorbital periods measured from 6 different instruments, including the one from Vela 5B observation reported by~\citet{pri84}. The horizontal lines indicate the durations of the corresponding light curves, and the dashed line represents the best fit of a linear model to estimate the period change rate, which yields a period derivative of $\dot P/P=(-7.13 \pm 0.33) \times 10^{-4}$ $yr^{-1}$ \label{lfsupper}}
\end{figure}

The power spectra are depicted in Figure~\ref{powspec}. Significant superorbital signals are evident in the power spectra of all six instruments. Additionally, the second harmonic signals are detectable in the data from the more sensitive instruments, including RXTE ASM, RXTE PCA, Swift BAT, and MAXI. The detected superorbital periods and their corresponding second harmonic periods are summarized in Table~\ref{supper}. It is apparent that the superorbital periods deviate from the period proposed by CG01, except for the one evaluated from Ginga ASM data, showing a tendency to decrease over time. By incorporating the superorbital period reported by~\citep{pri84} from the Vela 5B light curve, we estimated the timescale of the superorbital change by fitting a linear function to the detected superorbital periods over time (see Figure~\ref{lfsupper}). This result in a period derivative of $\dot P_{sup}/P_{sup}=(-7.13 \pm 0.33) \times 10^{-4}$ $yr^{-1}$, corresponding to an evolution timescale of 1,403 years. This period derivative exceeds the upper limit proposed by CG10 ($|\dot P_{sup} / P_{sup}| < 2.2 \times 10^{-4}$ $yr^{-1}$). However, the linear fitting yielded a reduced $\chi^2$ of  12.2, suggesting that the superorbital period evolution of 4U 1820-30 is likely more complex than the constant period derivative model suggests. Further variations in the superorbital period from phase analysis will be demonstrated in Section~\ref{pha}.

\begin{table*}[ht]
\centering
 \caption{Superorbital Period and RMS Amplitude Measured from the Light Curves Collected by Different Instruments \label{supper}}
\begin{tabular}{lcccccc}
\\
\tableline\tableline
Instrument  & Energy Rang & Duration & Superorbital& 2nd Harmodic & RMS AMP & RMS AMP\\
 & & &Period & Period & (CG01 ephemeris)& (Local ephemeris)\\
            &(keV)        & (MJD)    & (Days)  &  (Days)          & (\%) & (\%)\\
\tableline
Ginga ASM & 1-20 &46,861-48,532 & 170.95$\pm$1.16&  - & 33 & 36\\
RXTE ASM  &1.5-12 &50,083-55,927 & 169.03$\pm$0.06 &84.25$\pm$0.04 & 12 & 18\\
RXTE PCA  & 2-60 &51,215-55,864 & 168.06$\pm$0.09 & 84.04$\pm$0.06& 12 & 21\\
Swift BAT & 15-50 &53,415-60,157 & 167.62$\pm$0.10 &83.80$\pm$0.06& 7.6 & 15\\
MAXI      &2-20 &55,055-60,157 & 167.40$\pm$0.06 &83.06$\pm$0.04& 10 & 20\\
Fermi GBM &12-25 &55,967-60,157 & 166.62$\pm$0.34& - & 11 &15\\
\tableline
\end{tabular} 
\end{table*}

Additionally, to compare the amplitudes of superorbital modulation, we folded these  six light curves using two kinds of linear ephemeris. The first one is the optimal ephemeris proposed by CG01 (hereafter CG01 ephemeris),

\begin{eqnarray}\label{cg01eph}
 \nonumber  T_N &=& JD 2,450,909.9 + 171.033 \times N\\
                &=&MJD 50,909.4+171.033 \times N
\end{eqnarray}

\noindent  The other one is the local ephemeris, with a folding period corresponding to the best period obtained by the power spectrum (see Table~\ref{supper}), along with an arbitrary phase zero epoch for each light curve. The rms amplitudes folded by both types of ephemeris are listed in Table~\ref{supper}. The rms amplitudes of the profiles folded by the corresponding local ephemerides are significantly larger than those folded by CG01 ephemeris, indicating that the CG01 ephemeris is no longer suitable. This shows that the superorbital period of 4U 1820-30 has undergone significant changes during 1987 to 2023.

\subsection{Superorbital Phase Evolution}\label{pha}

In this research, we aimed to trace the long-term evolution of the superorbital phase of 4U 1820-30, necessitating the analysis of superorbital phases measured from different instruments. However, time lags between different energy bands are often observed in astronomical time series. For instance, soft phase lags of pulsed emissions are commonly noted in accreting millisecond X-ray pulsars~\citep{cui98,pat21}. Hence, a coherence test was conducted to verify if there was a significant time lag between any of two light curves from different instruments. However, this test could be only performed on the light curves with overlapping observation times. For each pair of light curves, only overlapping parts were selected for coherence test. The power spectra were obtained the superorbital periods for the corresponding light curves. The superorbital modulation profiles of both light curves were conducted by folding the mean period measured from the power spectra with an arbitrary but fixed phase zero epoch. We discovered that all the profiles could be well fitted with a four-component sinusoidal function, that is, $r(\phi)=a_0 + \sum_{k=1}^{4} [a_k cos(2\pi k \phi)+b_k sin(2\pi k \phi)]$. To measure the possible time delay between the two instruments, we applied the cross-correlation between the best-fitted modulation profiles. The test results are shown in Table~\ref{delay}. The phase difference is generally no more than 0.026 cycle, which is much smaller the phase jitters ($\sim$0.1 cycle, see CG01). We conclude that no significant systematic time delay is observable among these instruments for the superorbital modulation of 4U 1820-30.  

\begin{table}
\begin{center}
  \caption{Coherence Test for RXTE ASM, RXTE PCA, Swift BAT, MAXI and Fermi GBM Light Curves \label{delay}}

\begin{tabular}{llcr}
\\
  \tableline\tableline
Instrument 1 & Instrument 2 & Overlapping  Time   & Phase lag \\
           &     & (MJD)    & ($\phi_1 - \phi_2$)   \\
\tableline
RXTE ASM & RXTE PCA  & 51,215-55,864      & -0.009 \\
RXTE ASM & Swift BAT &  53,415-55,927      & 0.022  \\
RXTE ASM & MAXI &  55,055-55,927      &  -0.036 \\
RXTE PCA & SWIFT BAT & 53,415-55,864      &  0.022 \\
RXTE PCA & MAXI & 55,055-55,864      &  -0.007 \\
Swift BAT & MAXI&  55,055-60,157   & -0.021\\
Swift BAT & Fermi GBM &  55,967-60,157   & -0.001\\
MAXI & Fermi GBM &  55,967-60,157   & -0.009\\
\tableline
\end{tabular}
\end{center}
\end{table}

To trace the evolution of superorbital phase, we segmented the light curves and folded them to derive the modulation profiles. For instruments highly sensitive to superorbital modulation, like RXTE ASM and MAXI light curves, two cycles ($2 \times 171$ days) per segment sufficed to yield clear profiles. In the case of Swift BAT observations, where no superorbital modulation was detected for the photon energies higher than 24 keV~\citep{far09},  we selected four cycles as a data segment to ensure significant profile detection. As for the RXTE PCA light curve, due to the observation gaps, we adopted four cycles per segment to create the profiles. However, only three data segments provided sufficient phase coverage for further analysis. Given the very low sensitivity of Ginga ASM, a clear profile could only be obtained by folding the entire light curve.

\begin{figure}
   \epsscale{1.2}
\plotone{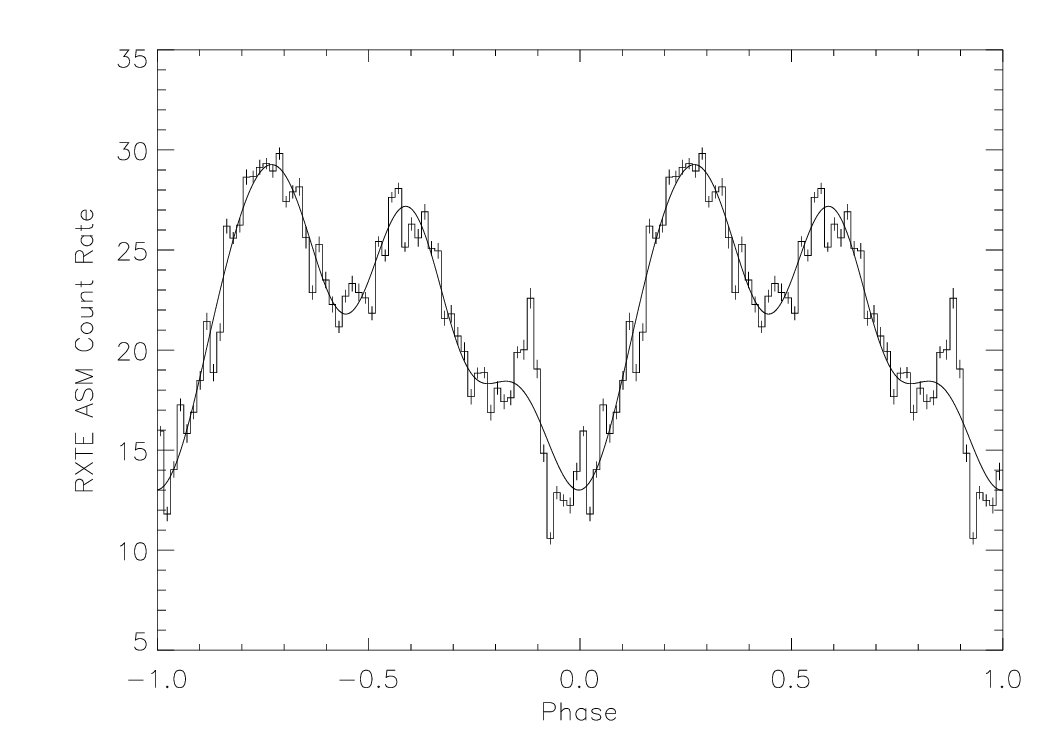}
\caption{A typical superorbital modulation profile of a data segment created by folding the light curve collected by RXTE ASM from MJD 54,158.64 to 54,497.86 ($\sim$2 cycles) with a folding period of 169.09 days from power spectral analysis (Section~\ref{ps}) and an arbitrary phase zero epoch. The solid line represents the optimal fit of a 4-component sinusoidal function to locate the superorbital minimum phase $\phi_{local}$. A BLS feature found by~\citet{sim03} can be also observed. \label{typprof}}
\end{figure}

Following the approach of  CG01, we selected the superorbital minimum as the fiducial point of the superorbital phase. Ideally , we would fold a light curve segment using a fixed ephemeris, such as the CG01 ephemeris (Eq.~\ref{cg01eph}) to determine the phase (i.e. $\phi_{CG01}$). However, as indicated in Section~\ref{ps}, the CG01 ephemeris is unlikely to be an optimal ephemeris for the all observations, particularly for recent ones (e.g. Swift BAT, MAXI, and Fermi GBM observations), which could lead to profile deformation. To precisely determine the $\phi_{CG01}$, we folded the light curve segments using the optimal linear ephemeris specific to each instrument (local ephemeris). This involved folding the data on the period obtained from power spectral analysis (see Table 1) and an arbitrary but fixed phase zero epoch. A typical modulation profile is depicted in Figure~\ref{typprof}. The phase ($\phi_{local}$) of a data segment was determined by fitting a four-component sinusoidal function and identifying the phase corresponding to the intensity minimum (fiducial point). This phase value, along with the local ephemeris, facilitated the evaluation of the superorbital minimum time ($t_m$) closest to the mid of observation time of the data segment. Subsequently, $t_m$ was then folded by CG01 ephemeris (Eq.~\ref{cg01eph}) to obtain the phase $\phi_{CG01}$.

The superorbital orbital phases ($\phi_{CG01}$) are listed in Table~\ref{phase}, and their  evolution is illustrated in Figure~\ref{phevol}. It is noteworthy that while the superorbital modulation displays strong periodicity in the power spectra, the modulation profile varies from cycle to cycle, exhibiting the quasi-periodic nature as described in ~\citet{zdz07a}. This variability induces phase jitters of $\sim$0.1 cycle, evident in Figure~\ref{phevol} and CG01. These phase jitters are considerably larger than the error estimated from photon statistics ($\sim$0.005 cycle). Despite the presence of phase jitters, a discernible phase evolution trend can be discerned in Figure~\ref{phevol}. However, independently evaluating errors from phase jitters is difficult, which depend on the evolution model. In the subsequent analysis, we utilized the unweighted fitting method outlined by~\citet{pre02} to update the ephemeris for the superorbital modulation of 4U 1820-30.

\begin{table}
\begin{center}
  \caption{Superorbital Phase ($\phi_{CG01}$) of 4U 1820-30\label{phase}}

  \begin{tabular}{cccl}
    \\
  \tableline\tableline
Superorbital minimum  & Phase      &      & Instrument \\
Time (MJD) & ($\phi_{CG01}$) &  &    \\
\tableline
47,677.560 &	0.1040  &   & Ginga ASM \\
50,240.528 & 	0.0892  &   & RXTE ASM \\
50,567.180 &	-0.0009 &   & RXTE ASM \\
50,886.301 &	-0.1351 &   & RXTE ASM \\
51,248.266 &	-0.0187 &   & RXTE ASM \\
51,617.490 &	0.1401  &   & RXTE ASM \\
51,943.259 &	0.0448  &   & RXTE ASM \\
52,272.795 &	-0.0285 &   & RXTE ASM \\
52,290.171 &	0.0755  &   & RXTE PCA \\
52,628.569 &	0.0517  &   & RXTE ASM \\
52,935.953 &	-0.1511 &   & RXTE ASM \\
53,278.853 &	-0.1462 &   & RXTE ASM \\
53,600.247 &	-0.2671 &   & RXTE ASM \\
53,789.674 &	-0.1595 &   & Swift BAT \\
53,967.300 &	-0.1210 &   & RXTE ASM \\
54,117.222 &	-0.2421 &   & RXTE PCA \\
54,299.651 &	-0.1778 &  & RXTE ASM \\
54,465.782 &	-0.2065 &  & Swift BAT \\
54,627.524 &	-0.2608 &  & RXTE ASM \\
54,974.529 &	-0.2319 &  & RXTE ASM \\
55,109.387 &	-0.4434 &  & Swift BAT \\
55,296.737 &	-0.3480 &  & RXTE ASM \\
55,132.435 &	-0.2946 &  & MAXI \\
55,467.179 &	-0.3491 &  & RXTE PCA \\
55,617.470 &	-0.4727 &  & RXTE ASM \\
55,794.143 &	-0.4257 &  & MAXI \\
55,802.064 &	-0.3934 &  & Swift BAT \\
55,980.165 &	-0.3381 &  & MAXI \\
56,472.080 &	-0.4760 &  & Swift BAT \\
56,478.810 &    -0.4226 &  & Fermi GBM\\
56,647.187 &	-0.4381 &  & MAXI \\
57,162.582 &	-0.4387 &  & Swift BAT \\
57,149.732 &    -0.5190  &  & Fermi GBM \\
57,154.762 &	-0.4704 &  & MAXI \\
57,481.630 &	-0.5593 &  & MAXI \\
57,817.333 &	-0.6105 &  & Swift BAT \\
58,135.365 &	-0.7370 &  & MAXI \\
58,279.709 &    -0.8931 &  & Fermi GBM \\	
58,450.326 &	-0.9095 &  & Swift BAT \\
58,483.492 &	-0.7016 &  & MAXI \\
58,815.675 &	-0.7594 &  & MAXI \\
58,975.301 &    -0.8261 &  & Fermi GBM \\
59,142.251 &	-0.8640 &  & Swift BAT \\
59,339.195 &	-0.6984 &  & MAXI \\
59,659.037 &    -0.8285 &  & Fermi GBM \\
59,805.990 &	-0.9832 &  & Swift BAT \\
59,939.781 &	-1.1869 &  & MAXI \\
60,149.720 &	-0.9594 &  & MAXI \\
  \tableline\tableline

\end{tabular}
\end{center}
\end{table}

\subsubsection{Linear Model}\label{linear}
\begin{figure}
   \epsscale{1.2}
\plotone{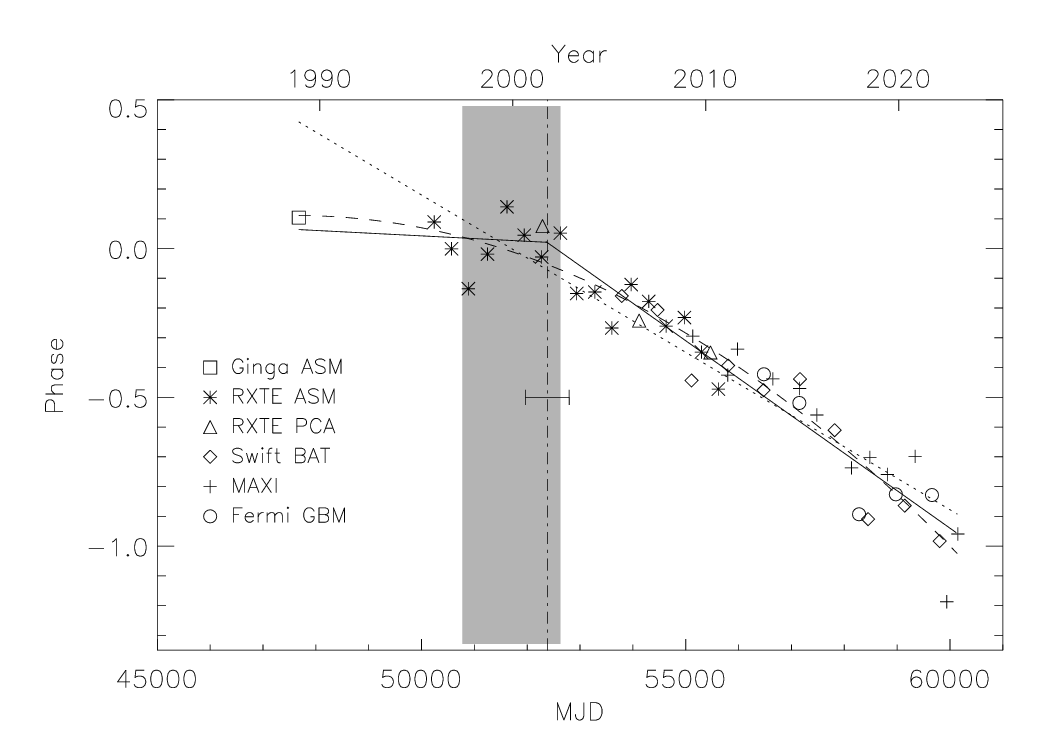}
\caption{Evolution of superorbital phases folded by the CG01 ephemeris from 1987 to 2023. The dotted, solid and dashed lines represent the best fits for linear, glitch and quadratic models, respectively. The shaded area indicates the low power state between MJD 50,773 and 52,627, and the vertical dash-dot line represents the glitch time MJD  $52,380 \pm 414$ evaluated by the glitch model, with the horizontal error bar indicating the 1$\sigma$ uncertainty of the glitch time. \label{phevol}}
\end{figure}

According to the triple model, the period should be remain stable from a long-term perspective because the superorbital modulation is induced by a hierarchical third component orbiting around the binary system. Therefore, our initial approach involved fitting a linear function to the phase evolution as depicted in Figure~\ref{phevol}. The parameters of the optimal linear function are listed in Table~\ref{para} yielding a period of $167.99 \pm 0.16$ days with a phase zero epoch of MJD $50,932.72 \pm 5.19$. We assessed the root-mean deviation (RMSD), defined as:
\begin{equation}\label{mdev}
 RMSD \equiv \sqrt{{1 \over \nu}{  \sum_{i=1}^N \bigl[{\phi_i - \phi(t_i)}\bigr]^2} }
\end{equation}

\noindent where $\phi_i$ is the detected phase, $\phi(t_i)$ is the expected phase value at $t_i$ evaluated from the best fit model, and $\nu$ is the degree of freedom. The RMSD is 0.12 for the linear model. However best-fitted period in this model significantly differs from the reported superorbital periods that  were detected in early RXTE ASM observations, as listed in Table~\ref{prep}, as well as the superorbital period of $176.4 \pm 1.3$ days reported by~\cite{pri84} from Vela 5B observation. Furthermore, the expected phase at the midpoint of Ginga ASM observation time (MJD 47,677.56) is $0.426\pm0.047$, about  6.9 $\sigma$ different from the detected value of 0.104 (see Figure~\ref{phevol}). Therefore, the linear model is unlikely to describe the superorbital phase evolution of 4U 1820-30.

\begin{table}
\begin{center}
  \caption{Parameters of Superorbital Modulation of 4U 1820-30\label{para}}

  \begin{tabular}{lll}
    \\
  \tableline\tableline
  \tableline\tableline
\multicolumn{3}{l}{Linear model} \\
\multicolumn{3}{l}{$\phi=a_0 + a_1  (t-T_0$)} \\
\multicolumn{3}{l}{$a_0= {{(T_0-T_{0,CG01})} / P_{CG01}}$\tablenotemark{a}} \\
\multicolumn{3}{l}{$a_1 = {{(P_0-P_{CG01})} / {(P P_{CG01})}}$} \\
\tableline
 Parameter& & Value \\
$a_0$ &  &$0.084 \pm 0.030$\\ 
$a_1 $ (cycle/day) & &$(-1.058 \pm 0.056) \times 10^{-4}$ \\ 
$cov(a_0 , a_1$) (cycle/day)& &$-1.43 \times 10^{-7}$\\
$T_0$ (MJD) & &$50,923.72 \pm 5.19$\\
$P$ (days)& &$167.99 \pm 0.16$\\

\tableline\tableline
\tableline\tableline
\multicolumn{3}{l}{Glitch model} \\
\multicolumn{3}{l}{$ \phi= \cases{a_0 + a_1  (t-T_0) & if $t \le T_g$; \cr a'_0 + a'_1  (t-T_0) & if $t > T_g$. \cr}$} \\
\multicolumn{3}{l}{$a_0= {{(T_0-T_{0,CG01})} / P_{CG01}}$} \\
\multicolumn{3}{l}{$a_1 = {{(P_1-P_{CG01})} / {(P_1 P_{CG01})}}$}\\
\multicolumn{3}{l}{$a'_0= {{(T_0-T_{0,CG01})} / P_{CG01}}+n_g (P_1 - P_2)/P_2$} \\
\multicolumn{3}{l}{$a'_1 = {{(P_2-P_{CG01})} / {(P_2 P_{CG01})}}$}\\
\multicolumn{3}{l}{$T_g = T_0 + n_g P_1 $}\\
\tableline
Parameter& & Value \\
$a_0$ &  &$0.034 \pm 0.027$\\ 
$a_1 $ (cycle/day) & &$(-9.12 \pm 19.29) \times 10^{-6}$ \\ 
$cov(a_0 , a_1$) (cycle/day)& &$ -8.47 \times 10^{-8}$\\
$a'_0$ &  &$0.205 \pm 0.043$\\ 
$a'_1 $ (cycle/day) & &$ (-1.260 \pm 0.071) \times 10^{-4}$ \\ 
$cov(a'_0 , a'_1$) (cycle/day)& &$ -2.82 \times 10^{-7}$\\
$T_0$ (MJD) & &$50,915.27 \pm 4.60$\\
$P_1$ (days)& &$170.77 \pm 0.56$\\
$P_2$ (days)& &$167.43 \pm 0.20$\\
$n_g$ & &$8.58 \pm 3.19$\\
$T_g$ (MJD) & & $52,379 \pm 414$\\
\tableline\tableline
\tableline\tableline
\multicolumn{3}{l}{Quadratic model} \\
\multicolumn{3}{l}{$\phi=a_0 + a_1(t-T_0)+a_2(t-T_0)^2$} \\
\multicolumn{3}{l}{$a_0= {{(T_0-T_{0,CG01})} / P_{CG01}}$} \\
\multicolumn{3}{l}{$a_1 = {{(P_0-P_{CG01})} / {(P_0 P_{CG01})}}$} \\
\multicolumn{3}{l}{$a_2 = 1/2$ $\dot P  / (P_0 P_{CG01})$} \\
\tableline
Parameter&  & Value \\
$a_0$ &  & $0.032 \pm 0.025$\\ 
$a_1 $ (cycle/day) & &$(-4.77 \pm 1.08) \times 10^{-5}$ \\ 
$a_2 $ (cycle/day$^2$) & &$(-7.24 \pm 1.24) \times 10^{-9}$ \\ 
$cov(a_0 , a_1$) (cycle/day)& &$-1.71 \times 10^{-7}$\\
$cov(a_0 , a_2$) (cycle/day$^2$) & &$1.10 \times 10^{-11}$\\
$cov(a_1 , a_2$) (cycle$^2$/day$^3)$& & $-1.24 \times 10^{-14}$\\
$T_0$ (MJD) & & $50,914.92 \pm 4.24$\\
$P_0$ (days) & & $169.65 \pm 0.31$\\
$\dot P$ (day/day) & & $(-4.20 \pm 0.72) \times 10^{-4}$\\
$\dot P / P_0$ (yr$^{-1}$) & & $(-9.04 \pm 1.55) \times 10^{-4}$\\ 
\tableline\tableline\tableline

\end{tabular}
\tablenotetext{a}{$T_{0,CG01}$=JD 2,450,909.9=MJD50,909.4 and $P_{CG01}$=171.033 days from Eq.9 of GC01.}
\end{center}
\end{table}

\subsubsection{Glitch Model}\label{glitch}
Table~\ref{prep} presents the reported superorbital periods detected by early RXTE ASM observations, which are roughly consistent with the period in the CG01 ephemeris (171 days). However, for later observations, particularly those from Swift BAT, MAXI, and Fermi GBM the period is approximately 167.4 days, as listed in Table~\ref{supper}. One the possibility is that the superorbital period underwent an abrupt change (glitch), likely between years 2000 and 2005 (see Figure~\ref{phevol}).

\begin{table}
\begin{center}
  \caption{Superorbital periods of 4U 1820-30 from Early RXTE ASM Observations \tablenotemark{a} \label{prep}}
  \begin{tabular}{cll}
    \\
\tableline\tableline
Observation Time & Period   & Reference \\
    (MJD)       &  (days)   &    \\
\tableline
50,088-51,606 & $171.39\pm 1.93$ \tablenotemark{b} & CG01\\
50,088-52,350 & 172.78  & \citet{sim03}\\
50,088-53,243 & $172 \pm 1$ & \citet{wen06}\\
50,088-54,151 & $170.6 \pm 0.3$ &\citet{zdz07a}\\
\tableline
\end{tabular}
\tablenotetext{a}{RXTE ASM was operated from MJD 50088 to 55831}
\tablenotetext{b}{The period of local ephemeris from Eq.8 in CG01.}
\end{center}
\end{table}

On the other hand,~\citet{kot12} conducted a dynamic power spectrum analysis and observed weaker superorbital modulation power during the period MJD 51,200-52,200, hereafter, referred to as the low power state. Additionally, signals with shorter periods, $\sim$85 days (second harmonic) and $\sim$65 days emerged in the dynamic power spectrum, indicating a change in the modulation profile during that time interval. However, considering the window size used to generate the dynamic power spectrum of 4U 1820-30 in~\citet{kot12} (5 cycles), this low power state time interval should be extended to approximately MJD 50,773 to 52,627. 

Compared with the phase evolution (i.e. Figure~\ref{phevol}), it appears likely that the glitch occurred around MJD 52,500, near the end of low power state. Marginal evidences support this assumption. As listed in Table~\ref{prep}, the reported superorbital period by \citet{zdz07a} was $170.6 \pm 0.3$ days, slightly smaller than those reported by CG01, \citet{sim03}, and \citet{wen06}. This discrepancy may be due to that a significant portion of data ($\sim 37 \%$) analyzed by~\citet{zdz07a} were collected after MJD 52,500. Similarly, the power spectral analysis of the entire RXTE ASM light curve yielded a superorbital period of 169.09 days (see Table~\ref{supper}), falling between 171 and 167.4 days, because about half ($\sim 56 \%$) of the data were collected after MJD 52,500. Conversely, the periods detected form the power spectra of Swift BAT, MAXI, and Fermi GBM were at approximately 167 days, because these observations were made after MJD 52,500 (see Table~\ref{supper}). 

\begin{figure}
   \epsscale{1.3}
\plotone{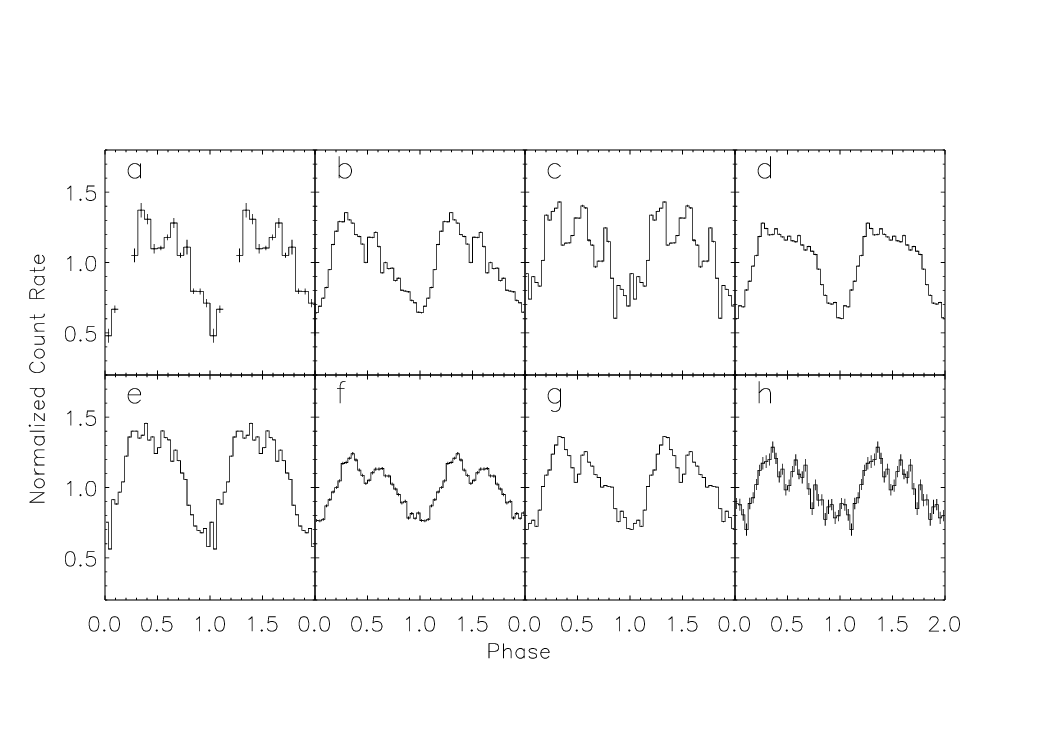}
\caption{Superorbital modulation profiles folded by glitch ephemeris listed in Table~\ref{para} for the light curves collected by (a) Ginga ASM, (b) RXTE ASM, before glitch, (c) RXTE PCA, before glitch, (d) RXTE ASM, after glitch, (e) RXTE PCA, after glitch, (f) Swift BAT, (g) MAXI, and (h) Fermi GBM. \label{mpglitch}}
\end{figure}

We therefore fitted the phase evolution with the glitch model using the ephemeris described by Eq.5 in~\citep{wol09}

\begin{equation}\label{gm}
  T_N = \cases{T_0 +P_1 N & if $N \le n_g$; \cr T_0 + P_1 n_g + P_2 (N-n_g)& if $N > n_g$. \cr}
\end{equation} 

\noindent where $T_0$ is the phase zero epoch, $P_1$ and $P_2$ are the periods before and after the glitch, respectively, $n_g$ is the glitch cycle count, and the glitch time $T_g \equiv T_0 + P_1 n_g$. The fitting results are shown in Figure~\ref{phevol}, and the parameters are listed in Table~\ref{para}. We obtained significantly different superorbital periods of $170.77 \pm 0.56$ days and $ 167.43 \pm 0.20$ days before and after the glitch time MJD 52,379$\pm$414, which occurred approximately between early 2001 and mid-2003, respectively. The change in the superorbital period is given by $\Delta P_{sup} /P_{sup}=-0.028 \pm 0.003$. The RMSD for this glitch model is 0.089. Comparing this with the RMSD of 0.12 from the linear model, the F-test yielded a p-value of 0.04, indicating that the glitch model is better than the linear model. Figure~\ref{mpglitch} shows the modulation profiles folded by the glitch ephemeris. All the superorbital minima (fiducial points) are close phase zero, implying that this ephemeris effectively describes the superorbital phase evolution of 4U 1820-30 from 1987 to 2023.

\subsubsection{Quadratic Model}\label{qua}
\begin{figure}
  \epsscale{1.2}
\plotone{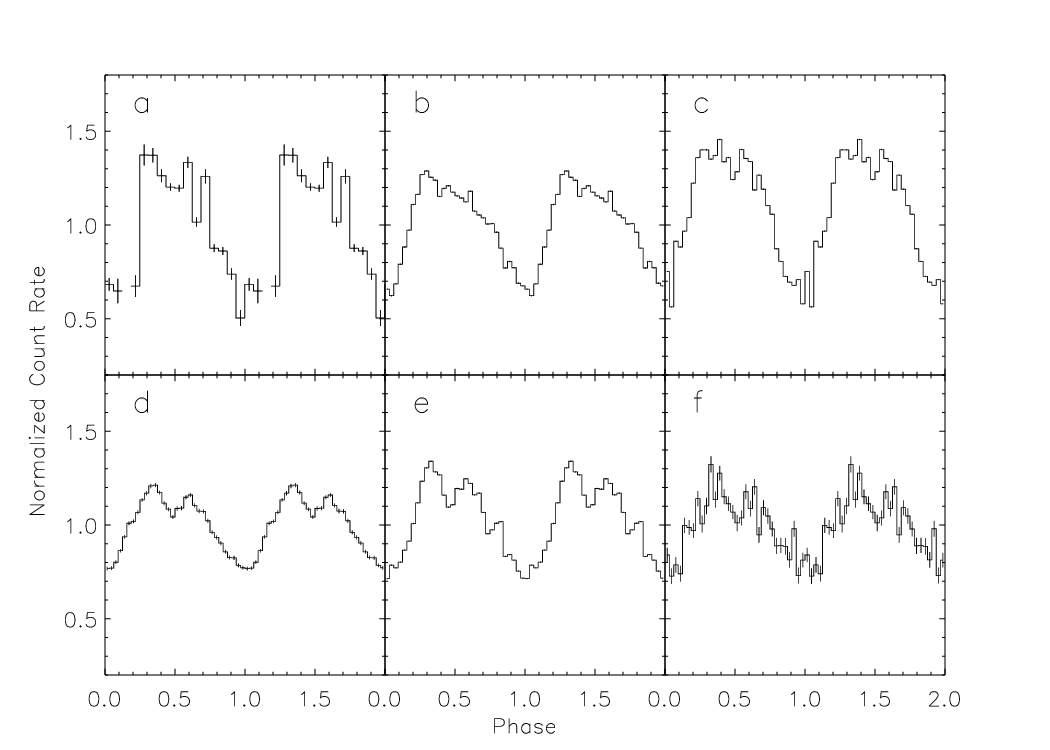}
\caption{Superorbital modulation profiles folded by quadratic ephemeris listed in Table~\ref{para} for the light curves collected by (a) Ginga ASM, (b) RXTE ASM, (c) RXTE PCA, (d), Swift BAT, (e) MAXI, and (f) Fermi GBM. \label{mpqua}}
\end{figure}

While the glitch model effectively describes the superorbital phase evolution, we cannot rule out the possibility that period difference between early RXTE ASM  observations and recent ones stems from a smooth change in the superorbital period.~\citet{far09} observed that the peak width of the superorbital signal in the power spectrum made from RXTE ASM and Swift BAT light curves were marginally wider than that from simulations, suggesting that a change in superorbital period. Hence, we apply a simple model assuming a constant period derivative ($\dot P_{sup}$) to fit a quadratic function to the superorbital phase evolution. The fitting results are depicted in Figure~\ref{phevol}, and the parameters are listed in Table~\ref{para}. A period derivative of $ \dot P_{sup}=(-4.20 \pm 0.72) \times 10^{-4}$ day/day, or $\dot P_{sup} / P_{sup}=(-9.04 \pm 1.55) \times 10^{-4} yr^{-1}$ was obtained from the fitting, and a quadratic ephemeris

\begin{eqnarray}\label{qeph}
  \nonumber  T_N &=& (MJD 50,914.92 \pm 4.24) +(169.65 \pm 0.31) \times N\\
                     &+&(-3.56 \pm 0.62) \times 10^{-2} \times N^2
\end{eqnarray}

\noindent was established. The RMSD for the quadratic  model is 0.087. Compared to this with the RMSD of 0.089 from glitch model, the F-test yielded a p-value of 0.55, which indicating that these two models are about equally adept at describing the superorbital phase evolution. Figure~\ref{mpqua} illustrates the modulation profiles folded by the quadratic model. Similar to the glitch model, all the superorbital minima (fiducial points) are located around phase zero. It provides an evidence that the quadratic model is suitable for describing the superorbital phase evolution of 4U 1820-30.

\subsection{X-ray Burst Active Times}\label{xbat}

As previously mentioned in Section~\ref{intro}, Type-I X-ray bursts of 4U 1820-30 are exclusively observed during the low state~\citep{cla77,ste84}. Further confirmed by CG01, indicated the Type-I X-ray bursts were detected only within $\pm$23 days around superorbital minima for bursts reported before 1985. This supports the hypothesis that the superorbital modulation stems from changes in the accretion rate changes rather than occultation effects. However, CG01's statistics only included four burst active dates. Subsequently, more X-ray bursts of 4U 1820-30 were detected. With the updated superorbital ephemerides, this evidence can be further substantiated. Although the possibility exists that the X-ray bursts occur in another low state, which may deviate significantly from the superorbital minima (e.g. the brief low state (BLS) found by~\citet{sim03}), it is likely that the most of X-ray burst active times would cluster around the superorbital minima. 

\begin{figure}
   \epsscale{1.2}
\plotone{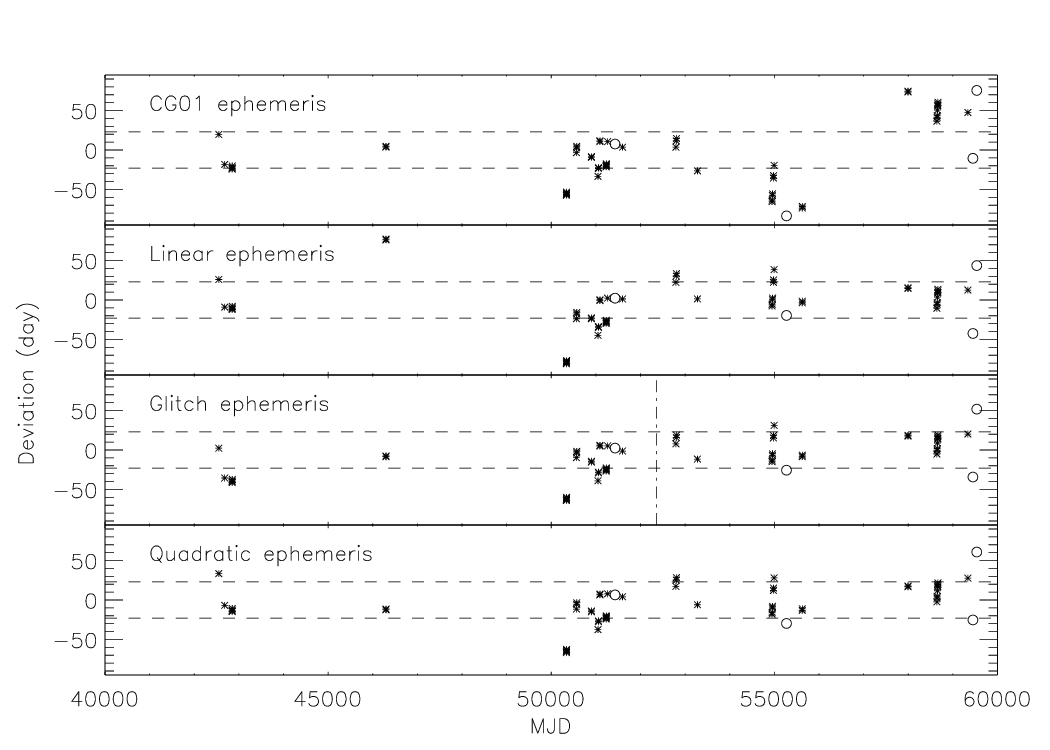}
\caption{The deviations of Type-I X-ray burst active dates relative to the nearest expected superorbital minima evaluated by CG01, linear, glitch and quadratic ephemerides.  The circles are the superbursts reported by~\citet{str02,int11,ser21a,ser21b}. The region between two horizontal dashed lines is the $\pm23$ days burst active interval suggested by CG01, and the vertical dash-dotted line in the plot of glitch ephemeris is the glitch time evaluated by the glitch model (Section~\ref{glitch}).\label{bph}}
\end{figure}

In this study, we collected the reported burst active dates of 4U 1820-30 from~\citet{gri76,vac86,hab87,zdz07b,gar13}, and~\citep{yu24}, as well as four superbursts discovered by~\citet{str02,int11,ser21a}, and~\citet{ser21b}. The deviations in burst active dates relative to the nearest superorbital minima predicted by CG01 and three ephemerides derived in this study are shown in Figure~\ref{bph}. It is evident that large deviations can be observed after the low power state (MJD $\sim$52,672) when using the superorbital minima predicted by the CG01 ephemeris. The root-mean-square deviations are 41.2, 33.5, 27.3 and 26.7 days for CG01, linear, glitch and quadratic ephemerides, respectively. This provides supportive evidence that glitch and quadratic ephemerides are better than CG01 and linear ephemerides in describing the superorbital phase evolution of 4U 1820-30.  Notably, X-ray bursts occurring between MJD 50,339 and MJD 50,343 exhibited large deviations ($\sim 60$ days) for all ephemerides. Upon examining the RXTE ASM light curve, we observed that these X-ray bursts occurred around a BLS with a count rate of only 12.0 cts/s compared to the mean count rate of 21.0 cts/s. This observation reaffirms that superorbital modulation is primarily caused by accretion variations rather than occultation effects.

On the contrary, out of the four superbursts detected, only the one on MJD 51,430~\citep{str02} fell within  $\pm$23 days region, whereas the other three occurred outside of this timeframe for both glitch and quadratic ephemerides (see Figure~\ref{bph}). Upon examining the RXTE ASM and MAXI light curves, we observed that the count rates on the dates of superbursts were 57\%, 150\%, 121\% and 138\% of the corresponding mean count rates of the light curves for the superbursts detected on MJD 51,430~\citep{str02}, MJD 55,272~\citep{int11}, MJD 59,449~\citep{ser21a} and MJD 59,543~\citep{ser21b}, respectively. It is likely that the low state constraint for the regular X-ray bursts of 4U 1820-30~\citep{cla77,ste84} does not apply to the superbursts. More observations are required for further confirmation.

\section{Discussion} \label{dis}
\subsection{Challenge of the Triple Model}\label{ctm}
The initial aim of this study was to further validate the stability of superorbital period of 4U 1820-30, a crucial piece of evidence for the triple model as described in Section~\ref{intro}, which explains its  superorbital modulation with a period of $\sim$170 days. Given that 4U 1820-30 resides in NGC 6624, a star-crowded region, previous studies suggested a high likelihood of the binary capturing a third star and forming a stable hierarchical triple system~\citep{gri88}.~\citet{bla82} proposed a stability criterion for such a triple system as

\begin{equation}\label{blc}
 \mu \leq \mu_{crit} ={ 0.175 {{\Delta^3} \over {(2-\Delta)^{3/2}}}}
\end{equation}

\noindent where $\mu=(m_2+m_3)/2m_1$, $\Delta=2(R-1)/(R+1)$, $R=R_3/R_1$, $m_1$, $m_2$ and $m_3$ are the masses of binary primary, secondary and tertiary companion, respectively, $R_1$ is the binary separation and $R_3$ is the maximum separation of the binary primary and the  tertiary companion. Applying Eq.~\ref{blc} to 4U 1820-30 system with the assumption that $m_1 = 1.4 M_{\sun}$, $m_2 = 0.07 M_{\sun}$~\citep{chou23}, $m_3 = 0.5 M_{\sun}$ (CG01), binary period of 685 s and third star orbital period of 1.1 days (see Section~\ref{intro}), according to the Kepler's third law, we found that $\mu = 0.204$ and $\mu_{crit} = 24.18$, which satisfies the  stability criterion proposed by~\citet{bla82}. Additionally, CG01, based on their analysis of the RXTE ASM light curve collected between 1996 and early 2000 and in conjunction with fiducial points measured by Vela 5B and Ginga, found no significant  superorbital period derivative, setting an upper limit of $|\dot P_{sup} / P_{sup} | < 2.2 \times 10^{-4}$ $yr^{-1}$, thereby confirming its stability and lending support to the triple model. Therefore, for the 4U 1820-30 system, with additional subsequent observations after 2000, one would expect that the observed superorbital period would closely match the value found by CG01 (171 days) and that the period derivative could be further constrained.

However, upon analyzing X-ray light curves collected by the sky monitoring/scanning instruments from 1987 to 2023, we discovered a significant change in the superorbital period from 171 days to 167 days, identified through both power spectral analysis (Section~\ref{ps}) and phase analysis (Section~\ref{pha}) over a time span of $\sim$36 years. This suggests that the ephemeris proposed by CG01 is no longer suitable for describing the superorbital modulation of 4U 1820-30, and the period is not as stable as anticipated by triple model. By analyzing the superorbital phase evolution, we suggested that the superorbital period may have experienced an abrupt change during  early 2001 to mid-2003 ($T_g= MJD 52,379 \pm 414$) or may be constantly changing with a period derivative of $\dot P = (-4.20 \pm 0.72) \times 10^{-4}$ day/day.

The significant difference between the period detected from Vela 5B observation, $176.4\pm 1.3$ days~\citep{pri84}, and Ginga observation, $171.12 \pm 1.99$ days (see Table~\ref{supper}) suggests that the superorbital period may have experienced another glitch between 1976 and 1987. If 4U 1820-30 is a hierarchical triple system, from Eq.~\ref{msp}, a glitch in superorbital period may be induced by changes in either the binary orbital period or the third star orbital period as

\begin{equation}\label{glip}
 {{\Delta P_{sup}} \over {P_{sup}}} = 2{{\Delta P_3} \over {P_{3}}} - {{\Delta P_{orb}} \over {P_{orb}}}
\end{equation}

\noindent Glitches in orbital periods have been observed in some LMXBs that  exhibit total eclipses, such as EXO 0748-676~\citep{wol09}, XTE J1710-281~\citep{jai11,jai22}, and AX J1745.6-2901~\citep{pon17}. These glitches likely result from magnetic, solar-type cycles of the companion star, affecting the mass distribution of companion and leading to variations in its quadrupole moment~\citep{wol09}. However, the magnitudes of these glitches are typically in the order of milliseconds, with $\Delta P_{orb} / P_{orb} \sim 10^{-7}-10^{-6}$. For 4U 1820-30 system, the superorbital period glitch was measured as $ |\Delta P_{sup} / P_{sup}|=2.8 \times 10^{-2}$. No glitch has ever been  observed in the binary orbital phase evolution~\citep[see Figure 4 in][]{chou23}, implying that the orbital period glitch of the third companion was as high as  $|\Delta P_3 / P_3| =1.4 \times 10^{-2}$, about 4 orders of magnitude larger than those from eclipsing LMXBs. However, the superorbital period change may not occur abruptly but within a finite short time interval. Suppose the timescale to be 800 days, estimated from the uncertainty of $T_g$, the mean orbital period derivative of the third star would be as high as $|\dot P_3| \approx 4 \times 10^{-3}$ day/day. Thus, the triple model is unlikely to explain this large superorbital period change in such a short time.

The low power state is a particular phase during the superorbital modulation evolution of 4U 1820-30. The dynamic power spectrum demonstrated in Figure 24 of~\citet{kot12} indicates that in addition to the weaker power detected in the superorbital period, the powers of its second harmonic and a signal of period $\sim$65 days became significant. This suggests that the modulation was more complicated than usual. Our phase analysis results also show that the superorbital phases had larger fluctuation during the low power state (see Figure~\ref{phevol}) with an RMSD of 0.11, evaluated by the best glitch model fitting, compared to 0.075 for the phases outside low power state. It is probable that the superorbital period was 171 days before the low power state, became unstable during the low power state, and stabilized at 167 days later. However, it is unclear what the cause of this phenomenon is.

\begin{figure}
    \epsscale{1.3}
\plotone{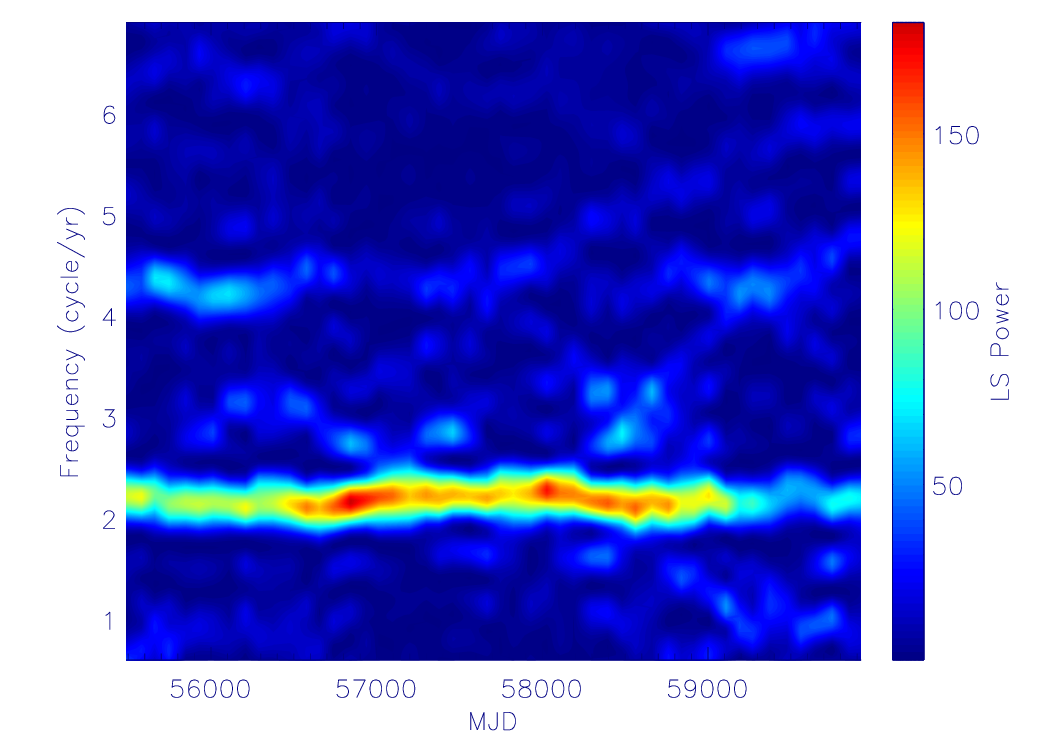}
\caption{The dynamic power spectrum generated from the light curve collected by MAXI. The power of superorbital modulation with a frequency$\approx$2.18 cycles/year was evidently lower after MJD $\sim$59,000. This implies that the 4U 1820-30 has entered a new low power state. \label{maxidps}}
\end{figure}

 Furthermore, from the phase evolution illustrated in Figure~\ref{phevol}, we observed that the phase fluctuation increased significantly after  MJD $\sim$58,000. The RMSD value, calculated at 0.13 using the best glitch model fitting, aligns with the value observed during the low power state. It suggested that 4U 1820-30 may be entering another low power state. To confirm this hypothesis, we generated a dynamic power spectrum of the light curve collected by MAXI as shown in Figure~\ref{maxidps}, using the same method as the one proposed by~\citet{kot12}. Notably, a decrease in power is evident after MJD $\sim$59,000. Considering the window size of five cycles, the onset of the new low power state is estimated to be around MJD 58,600. By comparing the start time of low power state observed in early RXTE ASM observation (MJD $\sim$50,773), the recurrent time of this phenomenon  is about 21.3 years. If the durations of these two low power states are roughly equivalent ($\sim$1,860 days), the new low power state is expected to end on about MJD 60,460, corresponding to 2024 May 30. Additional observations over the next a few years will enable us to confirm this prediction and verify the presence of a superorbital period glitch in this new low power state. Moreover, if the this phenomenon is periodically recurrent, the low power state preceding the one observed in early RXTE observation would likely have commenced around MJD 42,950, near the end of Vela 5B data period analyzed by~\citet{pri84} (mid of 1976), and end around MJD 44,800, before start time of Ginga ASM light curve (MJD 46,860). There may have been a superorbital period glitch causing the period change from 176.4 days to 171 days during this period.

Conversely, the superorbital phase evolution is also well-fitted by a quadratic model, although there is a significant difference (1.8$\sigma$) between the period  evaluated as ephemeris extrapolated to the midpoint of Vela 5B observation time ($173.54 \pm 0.97$ days) and the observed period ($176.4 \pm 1.3$ days) as reported by~\citet{pri84}. If 4U 1820-30 were a hierarchical triple system, from Eq.~\ref{msp}, the relation of period derivatives could be written as

\begin{equation}\label{quap}
 {{\dot P_{sup}} \over {P_{sup}}} = 2{{\dot P_3} \over {P_{3}}} - {{\dot P_{orb}} \over {P_{orb}}}
\end{equation}

\noindent Although the exact binary orbital period derivative is unknown, it is believed it is $\dot P_{orb} / P_{orb} \sim +10^{-7}$ yr$^{-1}$~\citep{chou23}. Therefore, the observed superorbital period derivative $\dot P_{sup} / P_{sup}= (-9.04 \pm 1.55)  \times 10^{-4}$ yr$^{-1}$ is contributed by the tertiary companion, with a value of $\dot P_{3} / P_{3}= -4.5 \times 10^{-4}$ yr$^{-1}$, corresponding to a variation timescale of only  $\sim$2,200 years. From Eq.~\ref{blc}, we infer that the triple system should be stable; therefore, such a fast period change is unlikely to occur. Furthermore, the acceleration from the gravitational potential in NGC 6624, estimated as $a_c /c \sim -10^{-7}$ yr$^{-1}$~\citep{peu14}, is insufficient for the orbital period derivative of the third companion derived from Eq.~\ref{quap}. Thus, the triple model can hardly explain the observed superorbital period derivative.


If 4U 1820-30 is not a triple system, the constraints regarding the observed value of the binary orbital period derivative may be relaxed. To explain the discrepancy between the positive theoretical value~\citep[$\dot P_{orb} / P_{orb} > 8.8 \times 10^{-8}$ yr$^{-1}$,][]{rap87} and the negative observed value~\citep[$\dot P_{orb} / P_{orb} =(-5.21 \pm 0.13) \times 10^{-8}$ yr$^{-1}$,][]{chou23} of the binary orbital period derivative, it has been proposed that 4U 1820-30 is being accelerated by the gravitational potential within the globular cluster NGC 6624~\citep{tan91,peu14}. However,~\citet{peu14} suggested that the maximum radial acceleration from the gravitational potential from NGC 6624 itself ($|a_{c,max}/c|=1.3 \times 10^{-9}$ yr$^{-1}$) is an order of magnitude smaller than the value required to explain the observed binary period period derivative. Therefore,~\citet{peu14} proposed  three possible scenarios to provide additional acceleration for 4U 1820-30, a flyby stellar mass dark remnant, a intermediate-mass black hole at the center of NGC 6624, and a central concentration of dark remnants. Only the last scenario was preferred because the first two scenarios tend to destroy the triple system~\citep{peu14}. However, if 4U 1820-30 is a pure binary system, the first two scenarios become viable explanations for the observed binary orbital period derivative.

\subsection{Thermal Disk Instability}\label{tdi}
The fact that the type-I X-ray bursts of 4U 1820-30 can only be observed in the low state~\citep{cla77,ste84} has been reconfirmed in this work (see Section~\ref{xbat}). This implies that the superorbital modulation of 4U 1820-30 is caused by variations in accretion rate rather than by external absorption or precession of accretion disk.~\citet{kot12} listed eight possible mechanisms to account for the superorbital modulations observed in X-ray binaries, but only the third body (i.e., triple model) and the X-ray state changes can possibly be responsible for the superorbital modulation of 4U 1820-30. If the triple model is ruled out due to the instability of superorbital period, the only remaining mechanism is the X-ray state changes. X-ray state changes refer to variations in mass accretion rate between high and low states due to thermal disk instability, as observed in dwarf novae and soft X-ray transients.~\citet{pri84} proposed that this mechanism could explain for the superorbital modulation of 4U 1820-30. However,~\citet{men02} pointed out that if thermal disk instability could occur in 4U 1820-30 system, the mass transfer rate ${\dot m} \leq {\dot m_{crit}=4.4 \times 10^{16}}$ g s$^{-1}$. From the mean flux of $<F_{bol}>=8.7 \times 10^{-9}$ erg cm$^{-2}$ s$^{-1}$ for 4U 1820-30~\citep{zdz07b} and the distance of  $8.019^{+0.108}_{-0.107}$ kpc for NGC 6624~\citep{bau21}, we obtained a mean luminosity of $<L>=6.7 \times 10^{37}$ erg s$^{-1}$ and a mass accretion rate of $\dot m_1 = 3.6 \times 10^{17}$ g s$^{-1}$ for a neutron star with a mass of 1.4 $M_{\sun}$ and a radius of $10^6$ cm. It is approximately an order of magnitude larger than the $\dot m_{crit}$. Therefore, this mechanism is unlikely to explain the superorbital modulation of 4U 1820-30.

\subsection{Irradiation-induced Mass Transfer Instability}\label{idi}
~\citet{zdz07b} discovered that the binary orbital modulation amplitude and the offset phase in 4U 1820-30 depend significantly on the accretion rate, which is highly related to the superorbital modulation phase. The orbital modulation in the X-ray band is believed to be caused by absorption from structures in the disk rim where the accretion flow from the companion impacts the outer edge of the disk~\citep{ste87}. As the mass loss rate changes, variations of the accretion stream induce changes in the absorption of outer edge structures and the position of impact point. This makes the amplitude and phase of orbital modulation dependent on the mass loss rate, and subsequently, on the accretion rate after a viscous time of $\sim 10^5$ s~\citep{zdz07b}. The variation in mass loss rate could be explained by the triple model as described in Section~\ref{intro}, where the eccentricity variation of the binary system induced by the third companion result in changes to the mass loss rate. Although the superorbital modulation is probably not a consequence of a third companion, the discovery by~\citet{zdz07b} implies that the superorbital modulation of 4U 1820-30 is due to changes in mass loss rate.   

One possible cause of variations in the accretion flow, aside from the presence of a third companion, is irradiation-induced mass transfer instability. This model has proposed to explain the flux variation of soft X-ray transients~\citep{ham86} and was included into the hybrid model proposed by~\citet{sim03}. Due to small binary separation of 4U 1820-30, the irradiation on the companion by the X-ray emission from the neutron star and the inner part of accretion disk is strong. Because the companion of 4U 1820-30 is only partially degenerate~\citep{rap87}, irradiation on the non-degenerate envelope enhances the mass loss of the companion.~\citet{chou23} estimated that at least 40\% of the mass lost from the companion is ejected from the binary system. Such a strong outflow is probably caused by the irradiation on the companion, as proposed by~\citet{tav91}. However, part of X-ray irradiation on the companion is blocked by the accretion disk, with the area that depending on the scale height of the disk rim. When the scale height of disk rim is small, a larger irradiation area enhances the mass loss rate and the accretion flow, which increases the scale height of the accretion disk rim. Conversely, when the scale height of accretion disk rim is large, a larger portion of the companion's surface is shielded by the disk. This results in a reduction in the mass loss rate, as well as in the accretion flow and the scale height of the accretion disk rim. This cyclical process may explain the quasi-periodic superorbital modulation of 4U 1820-30.

Suppose the accretion disk in 4U 1820-30 is geometrically thin and optically thick~\citep{pri81}. The shielded region on the companion can be estimated. For 4U 1820-30 with a neutron star mass of $m_1 =1.4 M_{\sun}$, a companion mass of $m_2 = 0.07 M_{\sun}$~\citep{chou23}, and an orbital period of $P_{orb} = 685$ s, we derived the binary separation of $a=1.33 \times 10^{10}$ cm from Kepler's third law. The discovery of the superhump in 4U 1820-30 system~\citep{wan10} indicates that the rim of the accretion disk reaches a 3:1 resonance radius, giving a disk radius of  $r_d = 6.4 \times 10^{9}$ cm. The scale height of the accretion disk can be evaluated as $\sqrt{kT/\mu m_H}/\Omega_k$~\citep{spr10} where $T$ and $\Omega_k$ are the temperature and the Keplerian angular velocity at the disk radius $r$, $m_H$ is the mass of a hydrogen atom, $\mu=4$ for a helium-dominated disk, and $k$ is the Boltzmann constant. The temperature at the disk rim is estimated as

\begin{equation}\label{tr}
 T={\Big({{3G \dot m_1  m_1} \over {8 \pi \sigma r_d^3}} \Big)^{1 \over 4}}
\end{equation}

\noindent \citep{pri81} where $\sigma$ is Stefan-Boltzmann constant and $G$ is gravitational constant. The accretion rate of 4U 1820-30 $\dot m_1 = 3.6 \times 10^{17}$ g s$^{-1}$. Thus, the temperature at the disk rim is $T=2.7 \times 10^4$ $K$ and the scale height is $H=2.8 \times 10^7$ cm. For a Roche lobe filling companion, the radius of companion is $R_2 =R_L = 2/3^{4/3} [q/(1+q)]^{1/3}a$~\citep{pac71} where $q=m_2/m_1$. For 4U 1820-30 system, q=0.05, so $R_2 =1.68 \times 10^9$ cm. The scale height of the irradiation shielded by the accretion disk on the companion around $L_1$ point is $h=H(a-R_2)/r_d =5.0 \times 10^7$ cm, which is equivalent to a latitude of $\sin ^{-1} (h/R_2)=1.7^{\circ}$ on the companion surface. Although the shielded latitude is small, it covers the $L_1$ point if the orbital plane and the accretion disk are coplanar. The accretion stream flows from a small region around the $L_1$ point on the surface of companion. If the accretion disk rim partially obscures this region, even a marginal change in irradiation on this region due to variations in the scale height of accretion disk rim could induce a significant change in mass loss because of the weak effective gravitational field around the $L_1$ point. Such a large variation in mass loss rate  could result in quasi-periodic superorbital modulation, causing a 2-3 fold change in the X-ray flux of 4U 1820-30. However, more observations and theoretical studies are required to verify this irradiation-induced mass transfer instability scenario, including the evolution of superorbital period discovered in this study.

\section{Conclusion} \label{con}

The triple model was once considered a plausible explanation for the superorbital modulation observed in 4U 1820-30. The stability of the superorbital period is the crucial evidence for verifying this model. CG01 suggested that the superorbital period was stable at 171 days and early RXTE ASM data support this 171-day periodicity, indicating stability of superorbital period.

In this study, we analyzed the data collected by Ginga ASM, RXTE ASM, RXTE PCA, Swift BAT, MAXI, and Fermi GBM over a time span of 36 years to verify the triple model for the 4U 1820-30 system. The superorbital periods derived from the power spectra of these six instruments show a significant change from 171 days to 167 days between 1987 and 2023, suggesting the instability of the superorbital period. Phase analysis revealed that the superorbital period may have experienced a period glitch between early 2001 and mid-2003, or may have changed smoothly with a period derivative of $\dot P_{sup}= (-4.20 \pm 0.72) \times 10^{-4}$ day/day. Two ephemerides, glitch and quadratic, were established to describe the expected superorbital minimum times of 4U 1820-30. These updated ephemerides accurately describe the superorbital minimum times with a mean phase jitters of $\sim0.08$ cycles. The fact that the Type-I X-ray bursts can be observed only in the low state implies a high probability of detecting the bursts around the superorbital minimum. By examining previously reported burst detection dates with different ephemerides, we found that the burst dates are more clustered around the superorbital phase zero when folded with the glitch and quadratic ephemerides, rather than with linear and CG01 ephemerides. This is not only reconfirms the low state constraint for regular X-ray bursts as suggested by~\citet{cla77} and~\citet{ste84}, but also provides supportive evidence that the glitch and quadratic ephemerides are better to describe the superorbital minimum times.

The instability of the superorbital periodicity in 4U 1820-30 discovered in this work seriously challenges the triple model. If the triple model were applicable to 4U 1820-30 system, according to Eq.~\ref{msp}, the superorbital period change could be due to either the binary period variation or the orbital period change of the third companion. However, the binary orbital modulation has been monitored for over 46 years, and neither a period glitch nor a large period derivative of ${\dot P}_{orb}/P_{orb} \sim 10^{-4}$ yr$^{-1}$ had previously been observed \citep[see][]{chou23}. Therefore, the superorbital period changes likely reflect the orbital period variation of the third companion. While orbital period glitches have been observed in some eclipsing LMXBs, the magnitude of these change is much smaller than that the superorbital period glitch observed 4U 1820-30. The period derivative derived from the quadratic model indicates that the timescale of the orbital period evolution of the third companion is $\sim 2200$ years, which is inconsistent with the expected stability in a hierarchical triple system. If the triple model does not apply to the 4U 1820-30 system, two previously unfavorable scenarios proposed by~\citet{peu14} - a stellar mass dark remnant and an intermediate mass black hole - can be reconsidered to explain the discrepancy between the theoretical and observed binary orbital period derivatives.

The absence of regular Type-I bursts in the high state suggests that the superorbital modulation of 4U 1820-30 results from variations in the accretion rate rather than from the occultation effect caused by the precession of a tilting or warping accretion disk. Thermal disk instability is unlikely to be the cause of the superorbital modulation due to the high accretion rate of 4U 1820-30. On the other hand, because the amplitude of orbital modulation highly depends on accretion rate~\citep{zdz07b}, the superorbital modulation could result from variation in mass transfer from the companion. Given the change in superorbital period of 4U 1820-30, such variation is unlikely to be induced by a third companion. We proposed that irradiation-induced mass transfer instability may be responsible for the superorbital modulation of 4U 1820-30. The accretion stream is expected to flow from a small region around the L$_1$ point on companion, where the effective gravitational field is weak. Therefore, the accretion stream is highly sensitive to the X-ray irradiation of this region. The irradiation onto this region may be partly blocked by the accretion disk rim, whose scale height also depends on the accretion stream. Small variations in the scale height can lead to  significant changes in accretion stream. A cyclical process could result in quasi-periodic superorbital modulation in 4U 1820-30.

Using the data collected by X-ray monitoring/scanning X-ray telescopes, we discovered the instability of the superorbital period of 4U 1820-30. From our study, we found that both the glitch model and the quadratic model describe the superorbital phase evolution well. However, additional observations are necessary to validate these models or to provide a better ephemeris for the superorbital modulation of 4U 1820-30. This period instability suggests that the triple model is unlikely suitable to explain the superorbital modulation of 4U 1820-30. Although we propose that the irradiation-induced mass transfer instability may be responsible for the superorbital modulation, further observations and theoretical works are required to verify this model, including the periodicity, modulation amplitude and profile, and the puzzling phase evolution, which were identified in this study. Fortunately, Swift BAT, MAXI and Fermi GBM continue to monitor the X-ray sky. Additionally, the newly operational Wide-field X-ray Telescope on-board the Einstein Probe~\citep{yua15}, which is sensitive to 0.5 to 4.0 keV X-ray photons and scans the entire night sky in three satellite orbits, can provide further data to better understand the nature of the superorbital modulation of 4U 1820-30.


\begin{acknowledgments}
The authors thank the anonymous referee for valuable suggestions that improved the manuscript. J.-L.W. and B.-C.C. especially acknowledge the support from the National Science and Technology Council of Taiwan through the grant NSTC 113-2112-M-008-002. This research has made use of data and software provided by the High Energy Astrophysics Science Archive Research Center (HEASARC), which is a service of the Astrophysics Science Division at NASA/GSFC. We also express our gratitude to the RXTE team for archiving the RXTE PCA monitoring observations of the galactic center and plane data, to Swift BAT transient monitor team for archiving the Swift BAT data, to MAXI team for archiving the MAXI data, and to Gamma-Ray Astrophysics Team of NSSTC for archiving the Fermi GBM data.
\end{acknowledgments}

%

\vspace{5mm}
\facilities{ADS, HEASRAC, Ginga (ASM), RXTE (ASM), RXTE (PCA),Swift (BAT), MAXI, Fermi (GBM)}



\software {heasoft(v6.31:\citet{hea14}), Astropy(\citet{apy13,apy18})}



\end{CJK*}

\end{document}